\documentclass[a4paper]{aa}
\usepackage{txfonts,graphicx,natbib}
\bibpunct{(}{)}{;}{a}{}{,}

\newcommand{\pv}{\ensuremath{P_V}}
\newcommand{\nnv}{\ensuremath{N_V}}
\newcommand{\geff}{\ensuremath{g_{\rm eff}}}
\newcommand{\bz}{\ensuremath{\langle B_z \rangle}}
\newcommand{\nz}{\ensuremath{\langle N_z \rangle}}
\newcommand{\cz}{\ensuremath{\langle c_z \rangle}}
\newcommand{\HeII}{He\,{\sc II}\,$\lambda4686$}

\begin{document}
\title{First constraints on the magnetic field strength\\ in extra-Galactic stars:\\
  FORS2 observations of Of?p stars in the Magellanic Clouds}
\author{
       S.~Bagnulo        \inst{1}
       \and
       Y.~Naz\'e\thanks{F.R.S.-FNRS Research Associate}\inst{2}
       \and 
       I.D.~Howarth        \inst{3}
       \and
       N.~Morrell        \inst{4}
       \and
       J.S.~Vink           \inst{1}
       \and
       G.A.~Wade         \inst{5}
       \and
       N.~Walborn        \inst{6}
       \and
       M.~Romaniello     \inst{7,8}
       \and
       R.~Barb\'a        \inst{9}
}
\institute{
           Armagh Observatory and Planetarium, College Hill, Armagh BT61 9DG, UK 
           \and
           Institut d'Astrophysique et de G\'eophysique, Universit\'e de Li\`ege,
           Quartier Agora (B5c), All\'ee du 6 Ao\^ut 19c, B-4000 Sart Tilman, Li\`ege, Belgium
           \and
           Department of Physics and Astronomy, University College London, Gower Street, London WC1E 6BT, UK
           \and
           Las Campanas Observatory, Carnegie Observatories, Casilla 601, La Serena, Chile
           \and
           Department of Physics, Royal Military College of Canada, P.O. Box 17000,
           Station Forces, ON, Canada K7K 4B4
           \and
           Space Telescope Science Institute, 3700 San Martin Drive, Baltimore, MD 21218, USA
           \and
           European Southern Observatory, Karl-Schwarzschild-Str.\ 2, D-85748,
           Garching bei M\"{u}nchen, Germany
           \and
           Excellence Cluster Universe, Garching bei M\"{u}nchen, Germany
           \and
            Departamento de F\'isica, Universidad de La Serena, Av. Juan Cisternas 1200 Norte, La Serena, Chile
}
\titlerunning{A magnetic survey in extra-Galactic stars}

\date{Received: 2016-11-04 / Accepted: 2017-02-24}

\abstract{
  Massive O-type stars play a dominant role in our Universe,
  but many of their properties remain poorly constrained. In the last
  decade magnetic fields have been detected in all Galactic members of
  the distinctive Of?p class, opening the door to a better knowledge
  of all O-type stars. With the aim of extending the study of magnetic
  massive stars to nearby galaxies, to better understand the role of
  metallicity in the formation of their magnetic fields and
  magnetospheres, and to broaden our knowledge of the role of magnetic
  fields in massive star evolution, we have carried out
  spectropolarimetry of five extra-Galactic Of?p stars, as well as a
  couple of dozen neighbouring stars.  We have been able to measure
  magnetic fields with typical error bars from 0.2 to 1.0\,kG,
  depending on the apparent magnitude and on weather conditions. No
  magnetic field has been firmly detected in any of our measurements,
  but we have been able to estimate upper limits to the field values
  of our target stars. One of our targets, 2dFS\,936, exhibited an
  unexpected strengthening of emission lines. We confirm the unusual
  behaviour of BI\,57, which exhibits a 787\,d period with two
  photometric peaks and one spectroscopic maximum.  The observed
  strengthening of the emission lines of 2dFS\,936, and the lack of
  detection of a strong magnetic field in a star with such strong
  emission lines is at odd with expectations. Together with the
  unusual periodic behaviour of BI\,57, it represents a challenge for
  the current models of Of?p stars.  The limited precision that we
  obtained in our field measurements (in most cases as a consequence
  of poor weather) has led to field-strength upper limits that are
  substantially larger than those typically measured in Galactic
  magnetic O stars. Further higher precision observations and
  monitoring are clearly required.
}

\maketitle
\section{Introduction}
About 5--10\,\% of Galactic OBA stars have detectable magnetic fields
\cite[e.g.,][]{Fosetal15,Wadetal16}. In these stars, the magnetic
field is generally associated with spectral peculiarities that result
from a variety of physical processes.

It has been known for a long time that most chemically peculiar A- and
B-type stars in the Galaxy rotate much more slowly than their
chemically normal counterparts in the same region of the HR diagram. A
large subset of these stars -- the Ap and Bp stars -- have strong
magnetic fields with a distinct characteristic: the observed field
strength changes with time, with the same period as the stellar
rotation (as deduced from photometric and spectroscopic measurements).
The explanation is given in terms of a stable magnetic field,
organised at a large scale, and not symmetric about the rotation axis
(e.g., a dipolar field with its axis of symmetry inclined relative to
the rotation axis), so that the observer sees a magnetic configuration
that changes as the star rotates \citep{Stibbs50}. Since the field
detection in the Galactic O stars $\theta^1$\,Ori\,C and
HD\,191612 by \citet{Donetal02,Donetal06}, it has become evident that
magnetic fields are also found in massive O-type stars, and modelling
such as that reported by \citet{Wadetal11} strongly suggests that the
magnetic fields of the slowly rotating O-type stars share the same
topological characteristics of Ap and Bp stars, i.e., their magnetic
field is dominated by a dipolar field tilted with respect to the
rotation axis. In this context, the distinctive category of Of?p stars
as defined by \citeauthor{Walborn72} (\citeyear{Walborn72}; see also
\citealt{Waletal11}) is of particular importance, as all known
Galactic Of?p stars have been found to be magnetic
\citep[e.g.,][]{Gruetal17}.

It is remarkable that stars ranging in mass from 1.5 to more than
$50\,M_\odot$ share such similar magnetic-field
characteristics. Despite this fact, the origin of the magnetic fields
in OBA-type stars is as yet uncertain. Whilst magnetic fields in
late-type stars are thought to be generated through dynamo action,
magnetic fields in OBA-type stars are likely of fossil origin
\citep[e.g.,][]{DonLan09}. Within the latter framework, different
  hypotheses have been put forward: conservation of the interstellar
  magnetic field trapped in the plasma during star formation and
  dynamos that acted during the earlier stages of pre-main sequence
  evolution \citep[both scenarios are reviewed by][]{Moss01}, and,
  more recently, mass transfer and mergers in close binary systems,
  either when two proto-stellar objects merge while approaching the
  main sequence and at least one of them has already acquired a
  radiative envelope \citep{Feretal09} or, at least for massive stars,
  during the main sequence \citep{Langer12}.

It is well understood that magnetic fields play a fundamental role in
stellar evolution by transporting angular momentum and affecting
stellar winds.  Since these processes depend sensitively on the opacities contributed by
metals, the conditions in low-metallicity dwarf galaxies might of
course be altogether different from those in a high metallicity spiral
such as our own Milky Way, but these aspects remain as yet completely
unexplored. Are these phenomena typical of our Galaxy only, or are
they found elsewhere? Is the field strength that characterises the
magnetic stars of our Galaxy also typical in nearby galaxies?
To the best of our knowledge, the physical effects of metallicity
  on frequency and strength of stellar magnetic fields in early-type
  stars have not been explored theoretically. Nevertheless, 
  investigating whether changes in metallicity have an impact on the
  formation and evolution of fossil fields would be very important.
  For instance, it has been
  recently proposed that during their life on the main sequence,
  massive stars develop an envelope inflation that is positively
  correlated with metallicity \citep[as a result of a change of the
    characteristics of the Fe opacity bump,][]{Graetal12,Sanetal17}. If
  confirmed, under the flux conservation one could speculate that
  fossil fields might be statistically stronger in massive stars with
  lower metallicity than in stars with higher metallicity. However, changes in metallicity
  may also alter the efficiency of interstellar medium flux advection
  (via changes to the electron density and ionization balance), or
  modify the characteristics of convection and rotation driving the
  dynamos of pre-main sequence stars (via changes to envelope
  opacities and the efficiency of wind/disc braking), making any theoretical
  prediction or even speculation particularly complicated. In conclusion,
  seeking guidance from observations provides an important motivation to search
  for and study extra-Galactic magnetic stars.

We know that the Magellanic
Clouds (MCs) host chemically peculiar stars \citep{Maietal01,Pauetal11}, but
they are too faint to be checked for magnetic fields with the
currently available instrumentation. However, there are five known
Of?p stars residing in the nearby MCs that are bright
enough to be within reach of today's instruments
\citep{Nazetal15,Waletal15}. The detection of their magnetic fields
through spectropolarimetric techniques would be tremendously exciting,
as these objects would be the first extra-Galactic stellar magnetic
fields to be directly discovered. Moreover, these objects are expected
to be different from their Galactic counterparts, as their metal
content ($Z$) is lower, and mass-loss rates are thought to be lower at
lower $Z$ \citep{Vinetal01,Moketal07}. One could hypothesise that
magnetic fields of massive stars in the MCs are
comparable to the Galactic ones because the observed spectroscopic
and photometric features variations are. On the other side, given that the Galactic Of?p
stars are thought to have a dynamical magnetosphere whose structure
depends on the capability of the magnetic field to channel and confine
the outflowing stellar wind, the similarities and differences in
Galactic and Magellanic Cloud Of?p stars would provide important
constraints about the interplay between stellar winds and magnetic
fields in low metallicity stars, providing a heretofore unavailable
chance to understand the role of magnetic fields in the earlier
Universe. 

In this paper we describe the results of a spectropolarimetric survey
of all Of?p stars known in the Large and Small Magellanic
Clouds. Specifically, we have searched for fields in three Of?p-type
stars in the SMC (SMC\,159-2, AzV\,220 and 2dFS\,936), and two
Of?p-type stars in the LMC (BI\,57 and LMC\,164-2). Taking advantage
of the multi-object capabilities of FORS2, we have also been able to
measure the magnetic field in several stars (typically 4-5) in the
close neighborhood of ($\la 3\arcmin$) of each main target.

\section{Observing strategy}
The targets that we selected for our survey had been identified as
Of?p stars because of their spectral peculiarities \citep[see][for a
  historical summary]{Waletal15}.  Recently, the analysis of
photometric datasets enabled us to detect the brightness variations of
our targets, deriving periods for four of them \citep{Nazetal15}.
\citet{Nazetal15} correlated the photometric variability of SMC\,159-2
to the spectral changes and \citet{Waletal15} did the same for
AzV\,220, BI\,57, and 2dFS\,936, demonstrating that the Of?p stars in
the Magellanic Clouds have similar behaviour to their Galactic counterparts.

\subsection{FORS2 spectropolarimetry}\label{Sect_FORS2_Spectropol}
To check whether these targets are magnetic, we obtained five half
nights of telescope time with the FORS2 instrument
\citep{AppRup92,Appetal98} at the ESO VLT. FORS2 is a multipurpose
instrument capable of imaging and low resolution spectroscopy, and
equipped with polarimetric optics (a retarder waveplate and a
Wollaston prism). For the measurement of the magnetic field we have
used the technique described by \citet{Bagetal02} and
\citet{Bagetal12}. The five extra-Galactic Of?p stars were
observed in multi-object mode, following the procedure already adopted
by \citet{Bagetal06} for a survey of magnetic stars in open clusters,
and by \citet{Nazetal12} for a magnetic survey of bright X-ray
emitters.

The detection of a typical magnetic field in such faint and hot stars is
just within the limits of the capabilities of the FORS2 instrument.
Using the results of \citet{Bagetal15} we predicted that we could
measure the longitudinal magnetic fields of our targets with a precision of
$\sim 250$\,G, which would allow us to reliably detect a field with
longitudinal component of $\sim 1$\,kG or higher. These predictions
were based on extrapolation through the relationship
\[
\sigma_{\bz} \propto \frac{1}{\mathrm S/N}
\]
applied to FORS1 archive data \citep[see Fig.~5 of][]{Bagetal15}.

Of course, chances to detect a magnetic field crucially depend not
only on the intrinsic strength of the magnetic fields of the targets,
but also on the geometrical view of the stellar field at the time of
the observations.

After the magnetic detections in Galactic O-type stars, it became
clear that the longitudinal field, UV/visible line profile changes (in
particular those of H$\alpha$), visible light curves of these stars
(when available), and X-ray emission strength are correlated, i.e. the
maximum brightness in visible and X-rays and the maximum emission of
the H$\alpha$ EW correspond to the maximum of the absolute value of the
longitudinal field. This is qualitatively explained as follows: the
field confines the wind towards equatorial regions, so when these
regions are seen face-on, the associated emissions are maximum while a
minimum occurs when they are seen edge-on
\citep{Sunetal12}. Generalizing this behaviour, we requested
new spectropolarimetric observations around the expected photometric
maximum, to maximise the probability of field detection. We note that
for the shorter period objects SMC\,159-2 and LMC\,164-2, the
ephemeris uncertainties yield uncertainties on the dates and phases of the photometric maxima
(at the time of our observations) of $\Delta\phi=0.07$ ($\Delta
t\sim$1\,d) and $\Delta\phi=0.05$ ($\Delta t\sim$0.4\,d),
respectively.  However, \textit{a posteriori}, the comparison of EW
measurements from our new FORS2 data with those previously obtained by
\citet{Nazetal15} and \citet{Waletal15} allowed us to check the phases
and assess whether our longitudinal field measurements were obtained
reasonably close to the emission maximum (see Sect.~\ref{Sect_EWs}) -
but note also that our scheduling requirements could not always be
met due to tight scheduling on the VLT.

\begin{figure}
\begin{center}
\scalebox{0.45}{
\includegraphics*[trim={0.8cm 4.7cm 0.3cm 2.8cm},clip]{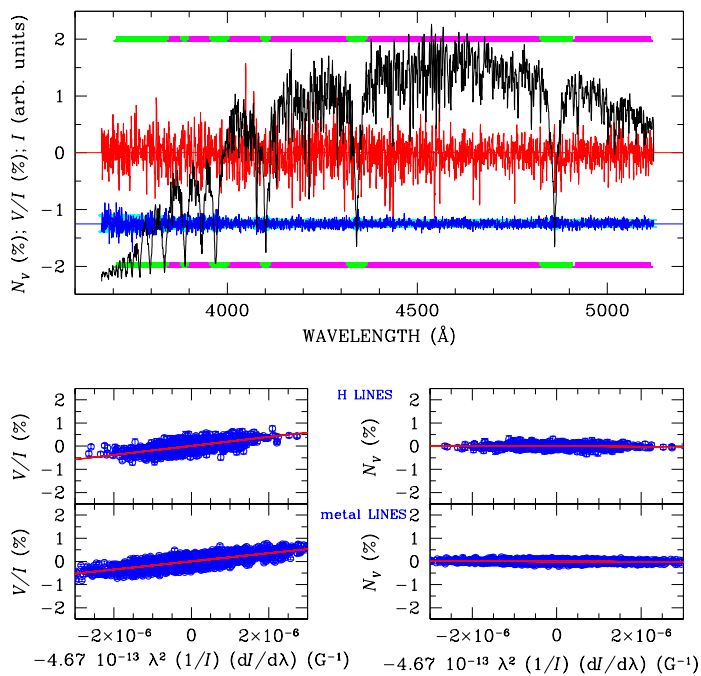}}
\caption{\label{Fig_HD188041a} FORS2 observations of the magnetic star
  HD\,188041.  In the upper panel, the black solid line shows the
  Stokes $I$ spectrum (uncorrected for the transmission function of the
  atmosphere + telescope and instrument optics); the red solid line
  shows the reduced Stokes $V$ spectrum, $\pv=V/I$ (in \% units), and
  the blue solid line is the null profile offset by $-1.25$\,\% for
  display purpose. The scattering of the null profile about zero is
  consistent (although sometimes slightly higher than) the 1\,$\sigma$ photon-noise error
  bars, which are also shown centred around
  $-1.25$\,\% and appear as a light blue background to the null
  profile.  Spectral regions highlighted by green bars (at the top and
  at the bottom of the panel) have been used to detemine the
  \bz\ value from H Balmer lines, while the magenta bars highlight the
  spectral regions used to estimate the magnetic field from He and
  metal lines. The four bottom panels show the best-fit obtained by
  minimising the $\chi^2$ expression of Eq.~(\ref{Eq_Chi}) using the \pv\ spectra (left
  panels) and the \nnv\ spectra (right panels) for H Balmer lines
  (upper panels) and metal lines (lower
  panels).}
\end{center}
\end{figure}
\begin{figure}
\begin{center}
\scalebox{0.45}{
\includegraphics*[trim={0.8cm 4.7cm 0.3cm 2.8cm},clip]{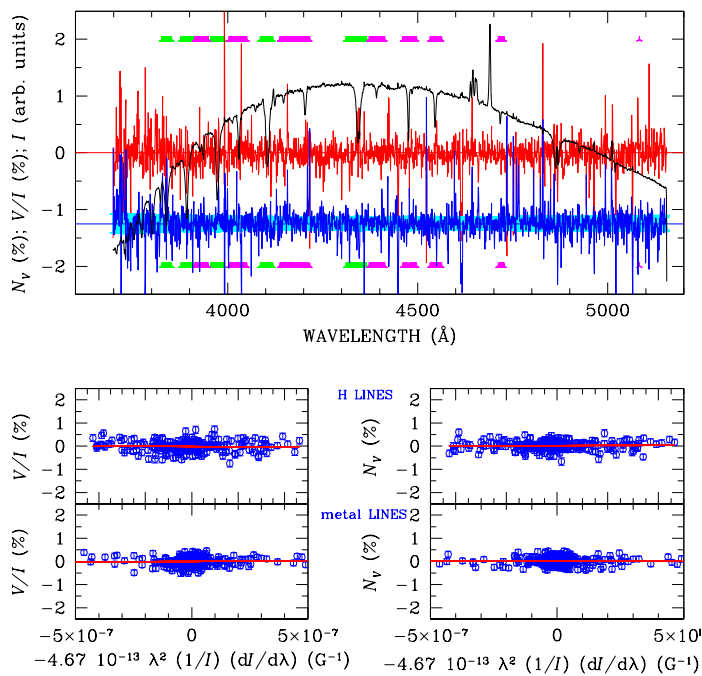}}
\caption{\label{Fig_LMC164} Same as Fig.~\ref{Fig_HD188041a} for
  LMC\,164-2 (observations obtained on 2016-01-05).}
\end{center}
\end{figure}
Observations of reference magnetic stars are not included in the
standard FORS2 calibration plan; nevertheless they are needed to confirm that
the position angle of the retarder waveplate is correctly reported by
the instrument encoders. For that reason, we decided to use some of
the twilight time to observe two well known and bright magnetic Ap
stars: HD\,94660, which has an almost constant longitudinal magnetic
field of $-$2\,kG \citep[e.g.][]{Lanetal14}, and HD\,188041, which has
a well known magnetic field that varies with a period of 223.78\,d
\citep{LanMat00}, and has been observed for more than 60 years,
starting from \citet{Babcock54}.

FORS2 is normally offered both in service and visitor mode with an MIT
CCD optimised for the red. In visitor mode it is possible to request
the use of the EEV CCD previously used in the now decommissioned FORS1
instrument, that is optimised for the blue. Since we were interested
in using the grism 1200B to cover the blue spectral region and since
our hot (and low reddened) targets emit more flux in the blue than in the red, for our
observing programme we requested the use of the EEV CCD. The actual
spectral range depends on the position of the MOS slitlet in the field
of view: with the slit in a central position, it was 3700--5120\,\AA.

\subsection{UVES-FLAMES spectroscopy}
H$\beta$ and H$\alpha$ observations of 2dFS\,936 were obtained on
2015-10-09 at UT\,06:58 (midpoint of 4.5\,h exposure) and 2015-10-10 at
UT\,03:34 (mid of a 3\,h exposure), on 2016-11-08 at UT 06:03
  (midpoint of a 4.3\,h exposure) and on 2016-11-09 at UT:01:23 (mid of a
  2.1\,h exposure) with the UVES spectrograph fed by FLAMES, using
the setting 580 which covers the spectral ranges 4790-5770\,\AA\ and
5840--6815\,\AA\ with a spectral resolution $\sim 50\,000$.

\section{Results}
\begin{figure*}
\begin{center}
\scalebox{0.60}{
\includegraphics*[angle=0,trim={0.7cm 0.7cm 0.3cm 0.8cm},clip]{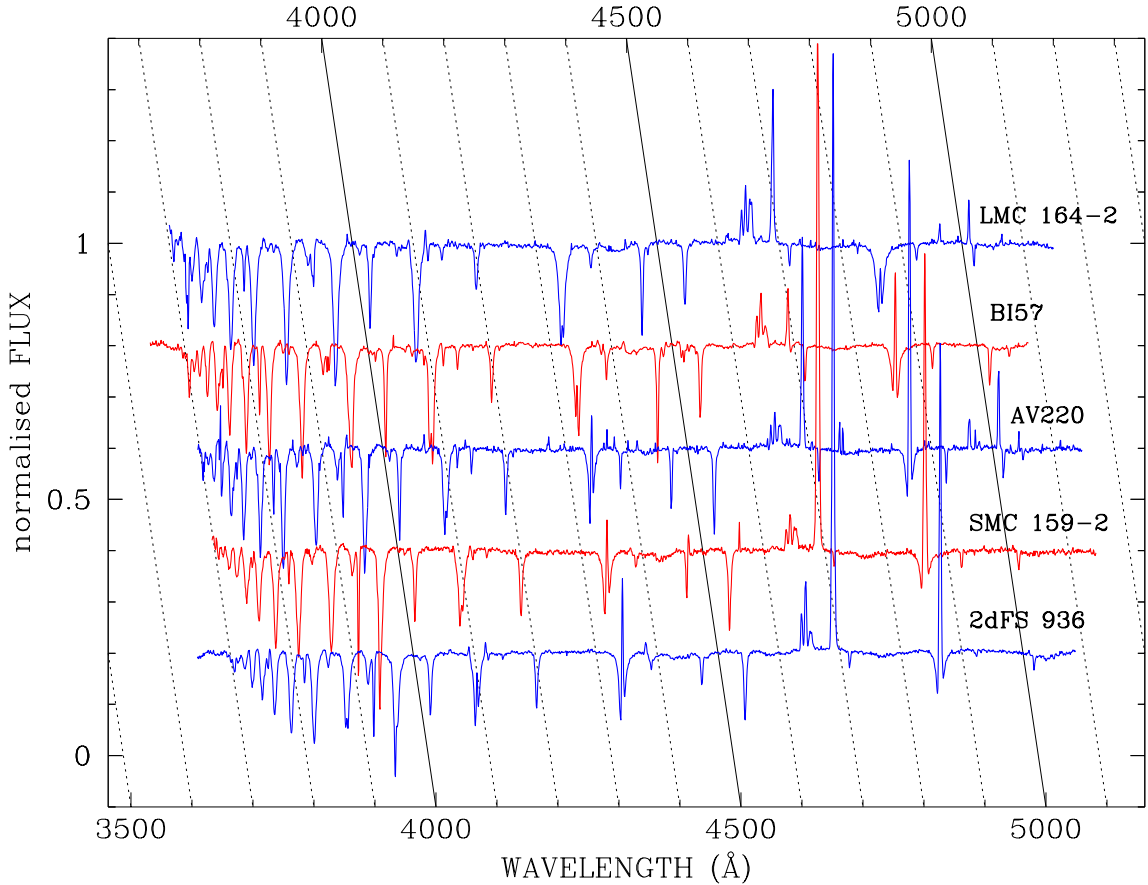}}
\caption{\label{Fig_I} Normalised FORS2 Stokes $I$ spectra of the five extra-Galactic O?fp stars.
For display purpose, spectra are offset both in $x$ and $y$ from each other.}
\end{center}
\end{figure*}

\begin{table*}
  \caption{\label{Tab_Log} Log of the FORS2 observations.  Columns~1
    and 2 give the civilian date and UT time of the midpoint of the
    observation; col.~3 gives the number of the slitlet where the star
    was located. Because of the Wollaston mask, only evenly-numbered
    slitlets are available for observations in spectropolarimetric
    mode; slitlets are numbered in increasing order from the top of
    the CCD to the bottom as described in the FORS user manual; the
    spectra of the star in slitlets from No. 2 to 10 are located on
    chip ``Norma'' and the spectra of the stars in slitlets from
    no. 12 to 18 on chip ``Marlene''. Part of the field of view of
    slitlet no.\ 12 falls on the gap between the two chips and in most
    of the cases could not be used. The main target was always placed
    in slitlet 10. Columns~4 and 5 give the J2000 RA and DEC 
    and the target
    name as identified through SIMBAD or other catalogues; col.~7 gives
    the $V$ magnitude, except entries flagged with a $*$ for which the
    magnitude refers to the $J$ filter; col.~8 is
    the star's spectral type; cols.~9 and 10 gives the total exposure
    time and the S/N per \AA; cols.~11 and 12 give our field
    determination from the reduced Stokes $V$ profiles, \bz, and from
    the null profiles, \nz.}
\begin{tiny}
\begin{tabular}{l@{\ \ \ }lcl@{\ \ \ }llrlrrr@{$\pm$}lr@{$\pm$}l}
\hline \hline
DATE  &                    
UT    &                    
N     &                    
\multicolumn{1}{c}{RA}&    
\multicolumn{1}{c}{DEC}&   
\multicolumn{1}{c}{STAR}&  
\multicolumn{1}{c}{$V$} &  
Sp. &                      
Exp &                      
S/N &                      
\multicolumn{2}{c}{\bz} &  
\multicolumn{2}{c}{\nz} \\ 
\hline

2015-10-08&23:48& 10 & 19:53:18.7 &$-$03:06:52&HD 188041               &$   5.6$&F0Vp     &   28 &3115 &$ 1780$&  26 &$  -80$&  10 \\ [2mm]

2015-10-09&05:59&  2 & 00:50:06.3 &$-$73:16:32& AzV 66                 &$  13.5$&B0I C    & 8400 &1195 &$  -70$& 185 &$  195$& 190 \\
2015-10-09&05:59&  4 & 00:50:17.5 &$-$73:17:18&2MASS J00501748-7317179 &$  15.1$&B0.5     & 8400 & 645 &$ -150$& 340 &$   25$& 360 \\ 
2015-10-09&05:59&  6 & 00:49:47.6 &$-$73:17:53& AzV 55                 &$  13.4$&B5I C    & 8400 &1335 &$  140$&  95 &$  -85$& 105 \\
2015-10-09&05:59&  8 & 00:49:59.7 &$-$73:18:42&2MASS J00495968-7318420 &$  15.2$&HPMS     & 8400 & 620 &$  335$& 340 &$  165$& 355 \\
2015-10-09&05:59& 10 & 00:49:58.7 &$-$73:19:28&{\bf SMC 159-2}         &$  15.1$&O8f?p    & 8400 & 550 &$ 2780$& 990 &$-2240$&1135 \\
2015-10-09&05:59& 14 & 00:50:04.8 &$-$73:21:03&2MASS J00500476-7321027 &$  15.0$&B0.5(V)  & 8400 & 590 &$ 1090$& 415 &$ -185$& 450 \\[2mm]

2015-10-09&08:27&  2 & 00:54:06.6 &$-$72:40:00&OGLE SMC-SC6 315697     &$  15.6$&B1-5     & 7200 & 435 &$ -415$&1140 &$  400$&1200 \\
2015-10-09&08:27&  4 & 00:54:02.3 &$-$72:42:22&OGLE SMC-SC6 311225     &$  15.2$&EB B0+B0.5&7200 & 790 &$ 1350$& 595 &$  -20$& 605 \\
2015-10-09&08:27&  8 & 00:53:42.2 &$-$72:42:35&AzV 148                 &$  14.1$&O8.5V    & 7200 &1230 &$    0$& 220 &$  225$& 215 \\
2015-10-09&08:27& 10 & 00:53:29.9 &$-$72:41:45&{\bf 2dFS\,936}         &$  14.1$&O6.5f?p  & 7200 &1405 &$ -965$& 530 &$-1120$& 540 \\
2015-10-09&08:27& 14 & 00:53:03.8 &$-$72:39:26&Dachs SMC 1-21          &$  13.6$&         & 7200 &1405 &$  230$& 220 &$  445$& 210 \\
2015-10-09&08:27& 18 & 00:52:52.5 &$-$72:44:13&SK 53                   &$  12.4$&B2Iab C  & 7200 &2130 &$   95$& 235 &$   20$& 230 \\[2mm] 

2015-10-10&02:43&  2 & 00:59:20.8 &$-$72:02:59&NGC 346 ELS 103         &$  16.2$&B0.5V    & 6000 & 280 &$   60$&1110 &$-3075$&1150 \\
2015-10-10&02:43&  4 & 00:59:20.8 &$-$72:03:38&NGC 346 ELS 100         &$  16.1$&B1.5V    & 6000 & 275 &$-1710$&1170 &$ 2475$&1200 \\
2015-10-10&02:43&  6 & 00:59:18.3 &$-$72:04:21&[BLK2010] flames1080    &$  16.1$&B0.5III  & 6000 & 345 &$ 2070$&1010 &$-1195$&1025 \\
2015-10-10&02:43&  8 & 00:59:04.2 &$-$72:04:49&NGC 346 ELS 68          &$  15.9$&B0V(Be-Fe)&6000 & 315 &$ 2450$&2550 &$-1370$&2600 \\
2015-10-10&02:43& 10 & 00:59:10.0 &$-$72:05:49&{\bf AzV 220}           &$  14.5$&O6.5f?p  & 6000 & 590 &$  515$& 575 &$-1695$& 670 \\
2015-10-10&02:43& 14 & 00:59:00.9 &$-$72:07:18&NGC 346 ELS 27          &$  15.0$&B0.5V    & 6000 & 360 &$-1545$& 930 &$  650$& 925 \\ 
2015-10-10&02:43& 16 & 00:59:05.6 &$-$72:08:02&NGC 346 ELS 19          &$  14.9$&A0II     & 6000 & 325 &$ 1135$&1300 &$ -135$&1235 \\ 
2015-10-10&02:43& 18 & 00:58:53.3 &$-$72:08:35&SkKM 179                &$  12.9$&K5V?     & 6000 & 345 &$ -480$& 580 &$  350$& 600 \\ [2mm] 

2016-01-05&02:43&  2 & 05:13:16.7 &$-$69:21:30& [M2002] LMC 92985=BI107&$  13.3$&B1:II    & 6000 &1395 &$-1440$& 580 &$  480$& 530 \\
2016-01-05&02:43&  4 & 05:13:26.7 &$-$69:21:55&  MACHO 5.5377.4508     &$  14.3$&B1:II    & 6000 &1200 &$ -605$& 325 &$  -85$& 320 \\
2016-01-05&02:43&  6 & 05:13:40.7 &$-$69:22:09& 2MASS 05134065-6922087 &14.7$^*$&B1:II    & 6000 &1265 &$  425$& 230 &$ -565$& 240 \\ 
2016-01-05&02:43&  8 & 05:13:38.8 &$-$69:23:00&2MASS 05133880-6922598  &13.8$^*$&Young SO & 6000 &1890 &$   85$& 160 &$ -235$& 160 \\
2016-01-05&02:43& 10 & 05:13:49.9 &$-$69:23:22& [MNM2014] {\bf LMC164-2}&$ 14.4$&O8f?p    & 6000 &1110 &$  205$& 560 &$  145$& 520 \\
2016-01-05&02:43& 18 & 05:14:25.6 &$-$69:25:02& SV* HV 2393            &$  15.0$&Class. Cep.&6000& 685 &$ -120$& 150 &$ -415$& 150 \\ [2mm]

2016-01-05&05:16& 10 & 10:55:01.0 &$-$42:15:04&HD 94660                &$   6.1$&Ap       &   80 &3610 &$-1893$&  21 &$    3$&  12 \\ [2mm]

2016-02-01&01:45&  2 & 05:01:36.9 &$-$68:08:59& 2MASS J05013694-6808585&14.9$^*$&         & 8640 & 200 &$ -545$&1120 &$-1615$&1125 \\
2016-02-01&01:45&  4 & 05:01:30.9 &$-$68:10:39& 2MASS J05013098-6810394&15.3$^*$&         & 8640 & 600 &$ -940$& 745 &$ 3170$& 800 \\ 
2016-02-01&01:45&  6 & 05:01:23.8 &$-$68:11:08& 2MASS J05012384-6811079&$14.9^*$&         & 8640 & 685 &$ 1410$& 620 &$  230$& 655 \\
2016-02-01&01:45&  8 & 05:01:14.9 &$-$68:10:44& 2MASS J05011491-6810440&$14.4  $& SRP     & 8640 & 265 &$ -670$&2160 &$-2590$&1960 \\
2016-02-01&01:45& 10 & 05:01:08.6 &$-$68:11:45& {\bf BI\,57}           &$14.0  $&         & 8640 &1335 &$ -360$& 345 &$ -440$& 330 \\
2016-02-01&01:45& 14 & 05:00:52.5 &$-$68:12:36& 2MASS J05005246-6812358&11.3$^*$&         & 8640 & 810 &$ -290$& 245 &$ -365$& 260 \\
2016-02-01&01:45& 16 & 05:00:47.8 &$-$68:13:57& 2MASS J05004675-6813567&15.8$^*$&         & 8640 & 625 &$    0$& 705 &$  965$& 770 \\
2016-02-01&01:45& 18 & 05:00:38.9 &$-$68:13:14& 2MASS J05003885-6813136&13.4$^*$&         & 8640 & 730 &$ -290$& 265 &$ -580$& 265 \\ [2mm]

2016-02-01&04:05&  4 & 05:13:13.3 &$-$69:19:56& 2MASS J05131332-6919555&$  14.5$&High PM  & 6000 & 660 &$ -345$& 290 &$ -200$& 280 \\
2016-02-01&04:05&  8 & 05:13:19.4 &$-$69:21:22& OGLE LMC-ECL-10254     &$  15.2$&ecl.var. & 6000 & 630 &$   -5$& 575 &$ -160$& 580 \\
2016-02-01&04:05& 10 & 05:13:26.7 &$-$69:21:55& MACHO 5.5377.4508      &$  14.3$& V*      & 6000 & 860 &$  375$& 365 &$ -435$& 330 \\ 
2016-02-01&04:05& 14 & 05:13:38.8 &$-$69:23:00& 2MASS 05133880-6922598 &13.8$^*$&Young SO & 6000 &1250 &$  -60$& 305 &$ -100$& 320 \\
2016-02-01&04:05& 16 & 05:13:49.9 &$-$69:23:22& [MNM2014] {\bf LMC164-2}&$ 14.4$&O8f?p    & 6000 & 670 &$ -550$& 870 &$ 1735$& 810 \\ [2mm]
2016-02-01&09:13& 10 & 10:55:01.0 &$-$42:15:04&HD 94660                &$   6.1$&Ap       &  120 &4445 &$-1885$&  20 &$  -12$&  10 \\ [2mm]
\hline
  \end{tabular}
\end{tiny}
\end{table*}

\begin{figure*}
\begin{center}
\scalebox{0.95}{
\includegraphics*[trim={1.6cm 5.9cm 1.3cm 9.2cm},clip]{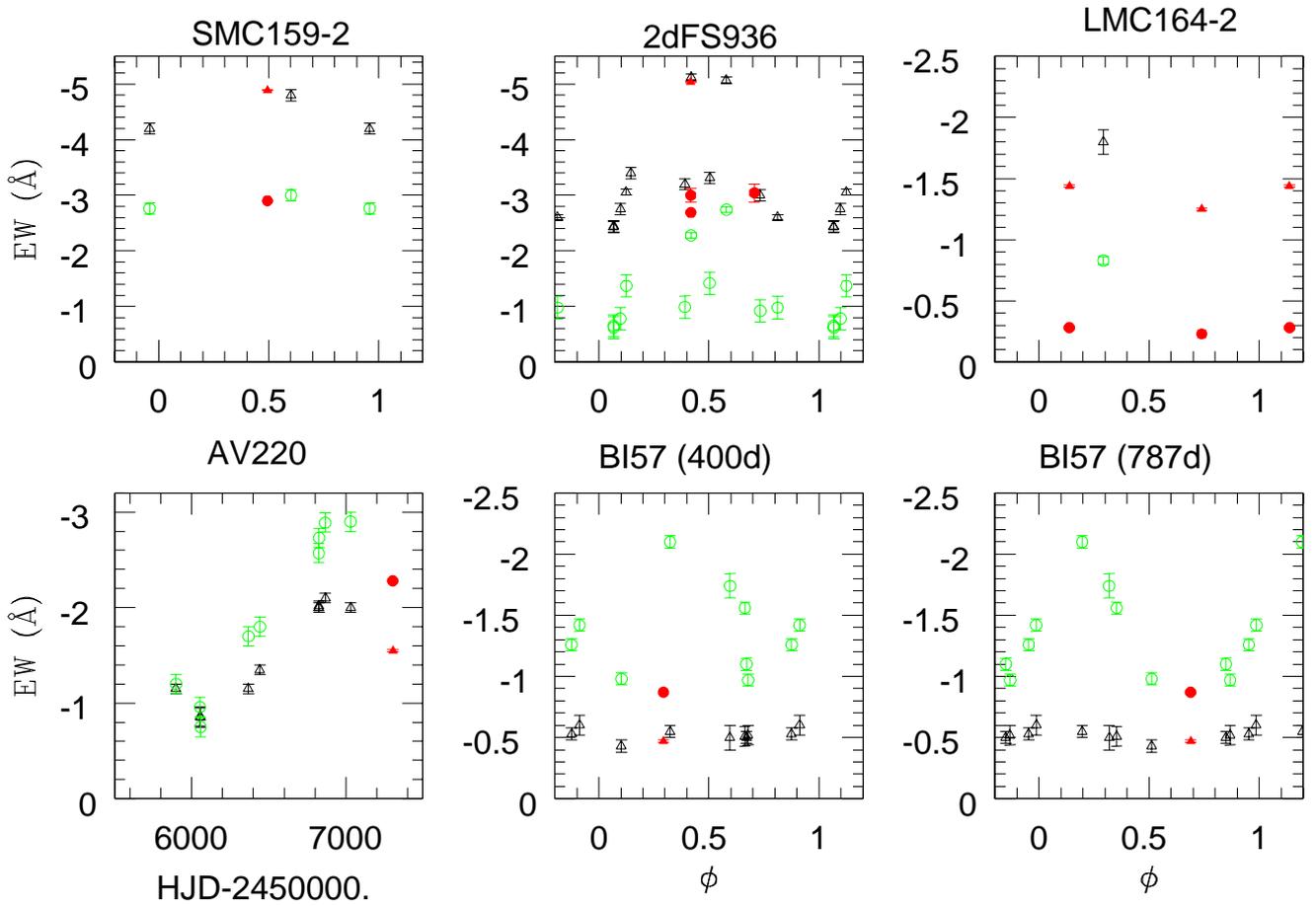}}
\caption{\label{Fig_EWs}
Measurements of the EW of the emission components for our five target
stars, plotted against Julian Date for AzV\,220 and against rotational
phase for the remaning targets \citep[ephemerides from][]{Nazetal15}.
For BI\,57 we have considered both the case of a rotation period of
400\,d and the case of a rotation period of 787\,d.
In all panels, triangles refer to \HeII, and circles to H$\beta$
measurements.  Red filled symbols refer to our new \HeII\ and H$\beta$
obtained with FORS2, FLAMES, and the B\&C given in Table~\ref{Tab_EWs}.
Empty symbols refer to data obtained in previous works, as detailed in
Sect.~\ref{Sect_EWs}.
}
\end{center}
\end{figure*}
\begin{figure}
\begin{center}
\scalebox{0.455}{
\includegraphics*[trim={1.3cm 4.8cm 1.1cm 2.8cm},clip]{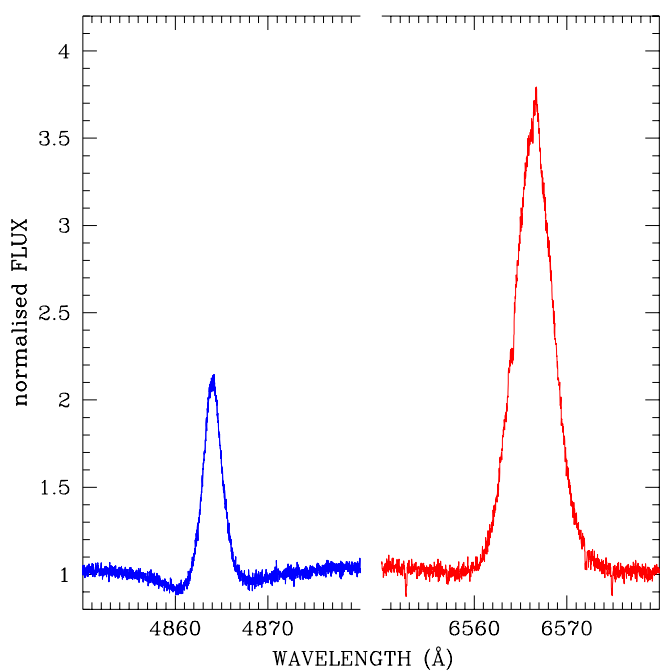}}
\caption{\label{Fig_Flames} H$\beta$ and H$\alpha$ of 2dFS\,936
  observed in October 2015 with UVES-FLAMES.}
\end{center}
\end{figure}
\begin{figure}
\begin{center}
\scalebox{0.455}{
\includegraphics*[trim={1.3cm 4.8cm 1.1cm 2.8cm},clip]{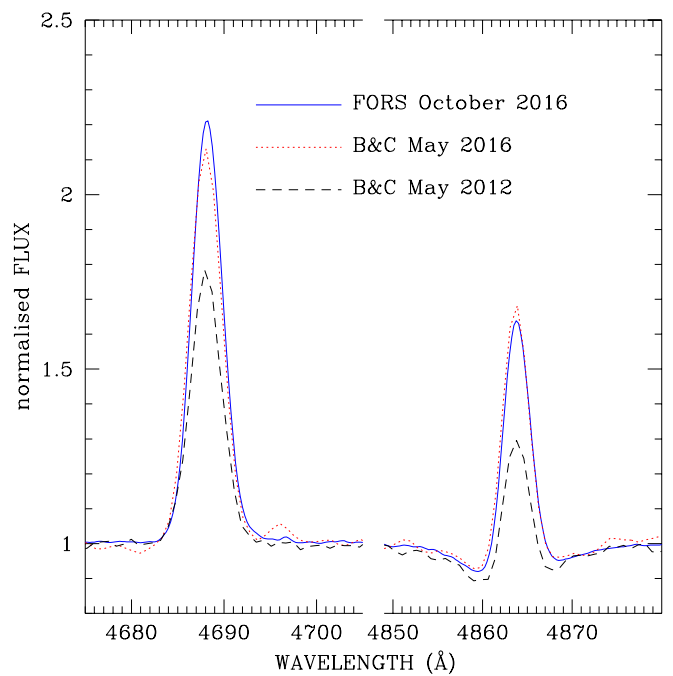}}
\caption{\label{Fig_2dFS} 2dFS\,936: \HeII\ and H$\beta$ of 2dFS\,936
  observed in May 2012 and in May 2016 with the B\&C spectrograph, and in October
  2015 with FORS2.}
\end{center}
\end{figure}

\begin{table*}
  \caption{\label{Tab_EWs} EWs of the HeII\,4686, H$\beta$ and,
    for 2dFS\,936 only H$\alpha$ lines.
   Column~3 indicates the phase using the ephemeris of Table~2 of
  \citet{Nazetal15} (for BI\,57 they refer to the period of 787\,d). Columns~4 to 6 show
  the total EWs obtained when integrating the full line profiles
  (i.e. absorption+emission; this was performed over the intervals
  4680--4700\AA\ for HeII\,4686, 4840--4890\AA\ for H$\beta$,
  and 6558-6574\,\AA\ for H\,$\alpha$). 
  Columns~7 to 9 give the EWs integrated over the emission component
  only considering as continuum the bottom of the line, as done by
  \citet{Waletal15}. }
  \begin{center}
    \begin{small}
  \begin{tabular}{lrrr@{\,$\pm$\,}lr@{\,$\pm$\,}lr@{\,$\pm$\,}lr@{\,$\pm$\,}lr@{\,$\pm$\,}lr@{\,$\pm$\,}l}
\hline
STAR    & HJD$-$& $\phi$ & \multicolumn{6}{c}{EWs (full profile)} & \multicolumn{6}{c}{EWs (emission only)} \\
        &2450000&        & \multicolumn{6}{c}{(\AA)}              & \multicolumn{6}{c}{(\AA)}               \\
        &       &        &\multicolumn{2}{c}{HeII4686}&\multicolumn{2}{c}{H$\beta$}&\multicolumn{2}{c}{H$\alpha$}&
                          \multicolumn{2}{c}{HeII4686}&\multicolumn{2}{c}{H$\beta$}&\multicolumn{2}{c}{H$\alpha$} \\
\hline
SMC\,159-2        & 7304.750 & 0.49 & $-4.90$ & 0.02     & $-1.61$ & 0.02 &\multicolumn{2}{c}{}& $-4.89$ & 0.01   & $-2.90$ & 0.01&\multicolumn{2}{c}{} \\
2dFS\,936 (FLAMES)& 7304.687 & 0.42 &\multicolumn{2}{c}{}& $-1.27$ & 0.59 &$-12.12$  &0.09& \multicolumn{2}{c}{}& $-3.00$ & 0.12&$-$12.12    & 0.09     \\
2dFS\,936 (FORS)  & 7304.853 & 0.42 & $-5.02$ & 0.02     & $-1.29$ & 0.02 &\multicolumn{2}{c}{}& $-5.04$ & 0.01   & $-2.69$ & 0.01&\multicolumn{2}{c}{} \\
2dFS\,936 (B\&C)  & 7526.923 & 0.58 & $-4.83$ & 0.08     & $-1.51$ & 0.012&\multicolumn{2}{c}{}& $-5.07$ & 0.06   & $-2.74$ & 0.05&\multicolumn{2}{c}{} \\
2dFS\,936 (FLAMES)& 7701.155 & 0.71 &\multicolumn{2}{c}{}& $-2.44$ & 1.04 &$-12.45$  &0.46& \multicolumn{2}{c}{}& $-2.98$ & 0.09&$-$12.45    & 0.46     \\
AzV\,220          & 7305.614 &      & $-1.56$ & 0.02     & $-0.83$ & 0.10 &\multicolumn{2}{c}{}& $-1.55$ & 0.01   & $-2.28$ & 0.01&\multicolumn{2}{c}{} \\
BI\,57            & 7419.572 & 0.69 & $-0.28$ & 0.01     & $ 0.77$ & 0.10 &\multicolumn{2}{c}{}& $-0.47$ & 0.01   & $-0.87$ & 0.01&\multicolumn{2}{c}{} \\
LMC\,164-2        & 7392.613 & 0.74 & $-1.25$ & 0.03     & $ 1.92$ & 0.10 &\multicolumn{2}{c}{}& $-1.25$ & 0.01   & $-0.23$ & 0.01&\multicolumn{2}{c}{} \\
LMC\,164-2        & 7419.669 & 0.14 & $-1.42$ & 0.02     & $ 1.98$ & 0.10 &\multicolumn{2}{c}{}& $-1.44$ & 0.01   & $-0.28$ & 0.01&\multicolumn{2}{c}{} \\
\hline \hline
  \end{tabular}
  \end{small}
\end{center}
\end{table*}

\subsection{Longitudinal magnetic field measurements}
Data were reduced as explained by \citet{Bagetal15}. The
mean longitudinal magnetic field \bz\ (i.e., the component of the
magnetic field averaged over the visible stellar disk)
was calculated by minimising
the expression
\begin{equation}
\chi^2 = \sum_i \frac{(y_i - \bz\,x_i - b)^2}{\sigma^2_i}
\label{Eq_Chi}
\end{equation}
where, for each spectral point $i$, $y_i = \pv(\lambda_i)$, $x_i =
-g_\mathrm{eff} \cz \lambda^2_i (1/I_i\ \times
\mathrm{d}I/\mathrm{d}\lambda)_i$, and $b$ is a constant introduced to
account for possible spurious polarisation in the continuum. As a
quality check, field measurements were also estimated from the null
profiles \citep[see][for an extensive discussion on the use of null
  profiles for quality check]{Bagetal12}.

For the field measurement we considered three cases: in Eq.~(\ref{Eq_Chi}) we first
  used the spectral points including H Balmer lines only
\citep[adopting $\geff=1$ for the Land\'e
  factor,][]{CasLan94}, then we included He and metal lines only (setting \geff=1.25),
and finally we included all (H, He and metal) spectral lines together. We also
avoided emission lines to be sure to probe the stellar photosphere
rather than the circumstellar environment. As the
results of these three measurement procedures roughly agree, we
report here only the last value, which also yields the smallest error
bars.  Figure~\ref{Fig_HD188041a} shows an example of field detection
on one of the magnetic reference stars that we observed to check the correct
alignment of the polarimetic optics.  Figure~\ref{Fig_LMC164} shows
the same plots for the science target Of?p target LMC\,164-2.  Our
full list of measurements is given in Table~\ref{Tab_Log}. 

\subsection{Equivalent width measurements}\label{Sect_EWs}
FORS2 Stokes $I$ spectra of all extra-Galactic Of?p stars are shown in
Fig.~\ref{Fig_I}. From these spectra we have measured the EWs of the
\HeII\, and H$\beta$ lines. For the star 2dFS\,936 we have also
  measured the EW of \HeII\ and H$\alpha$ from the UVES-FLAMES spectra
  (obtained in October 2015 and November 2016), and the EW of
  \HeII\ and H$\beta$ from the spectra obtained in May 2016 with the
  Boller \& Chivens (B\&C) spectrograph of the du Pont telescope at
  Las Campanas Observatory.  All these new measurements 
  are reported in Table~\ref{Tab_EWs}, and shown in Fig.~\ref{Fig_EWs}
  together with EW measurements obtained previously by
  \citet{Waletal15} (for AzV\,220, 2dFS\,936 and BI57), by
  \citet{MasDuf01} (for 2dFS\,936), and by \citet{Masetal14} (for
  SMC\,159-2 and LMC\,164-2).  In the following we comment on
  individual stars.

\begin{figure}
\begin{center}
\scalebox{0.45}{
\includegraphics*[trim={0.7cm 5.7cm 0.3cm 2.8cm},clip]{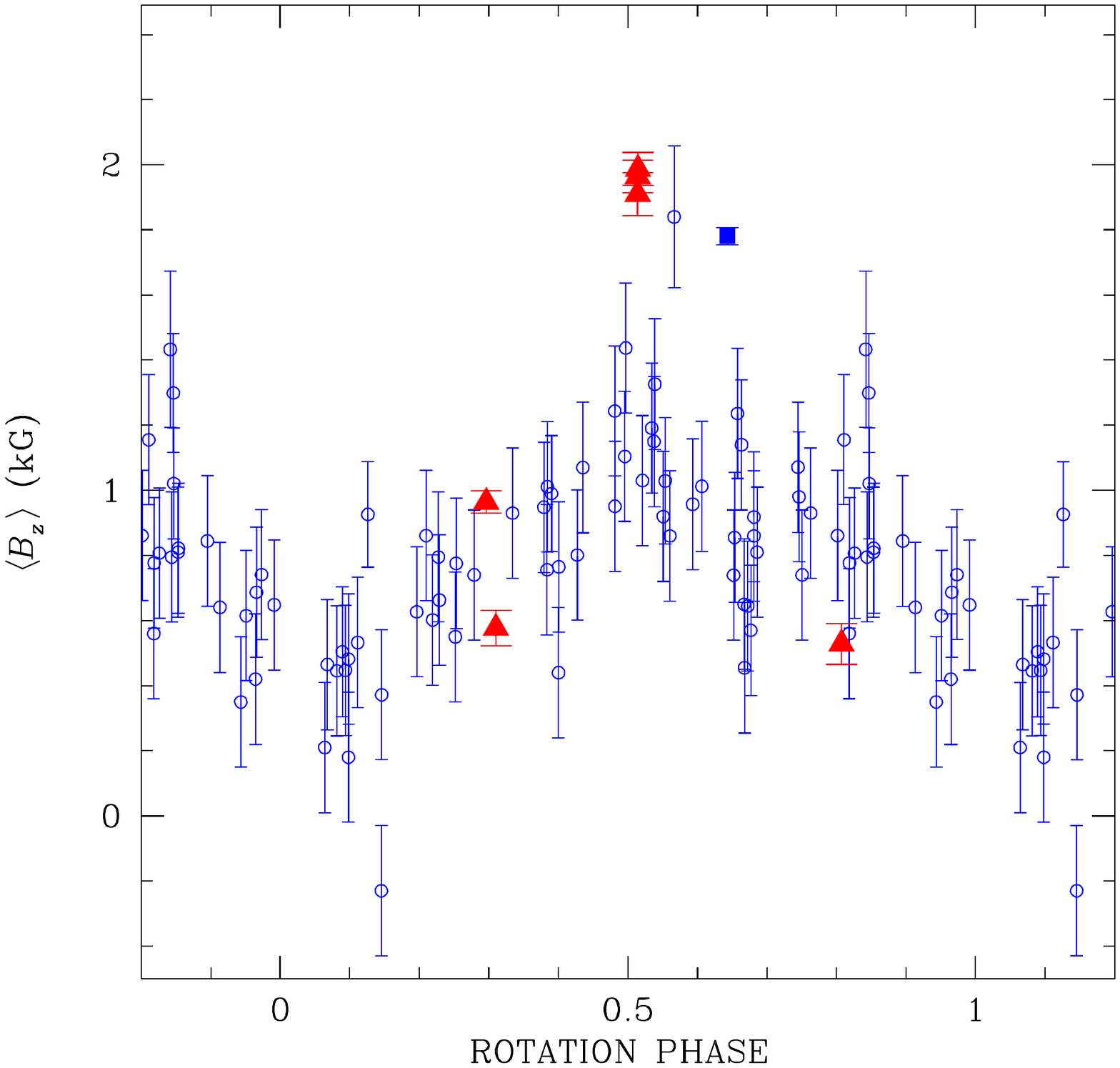}}
\caption{\label{Fig_HD188041b} Longitudinal field measurements of the magnetic
  star HD\,188041. Blue empty circles are from previous works
  \citep{Babcock54,Babcock58,Mathys94,MatHub97} and the blue solid square
  from this work. Measurements have been phased with the rotation period
  of 224.78\,d \citep{LanMat00}. Red solid triangles refer to previous
  field measurement obtained with FORS1 and with different grisms
  \citep[see][and references therein]{Lanetal14,Bagetal15}.}
\end{center}
\end{figure}
\begin{figure}
\begin{center}
\scalebox{0.45}{
\includegraphics*[trim={0.8cm 0.8cm 0.3cm 0.8cm},clip]{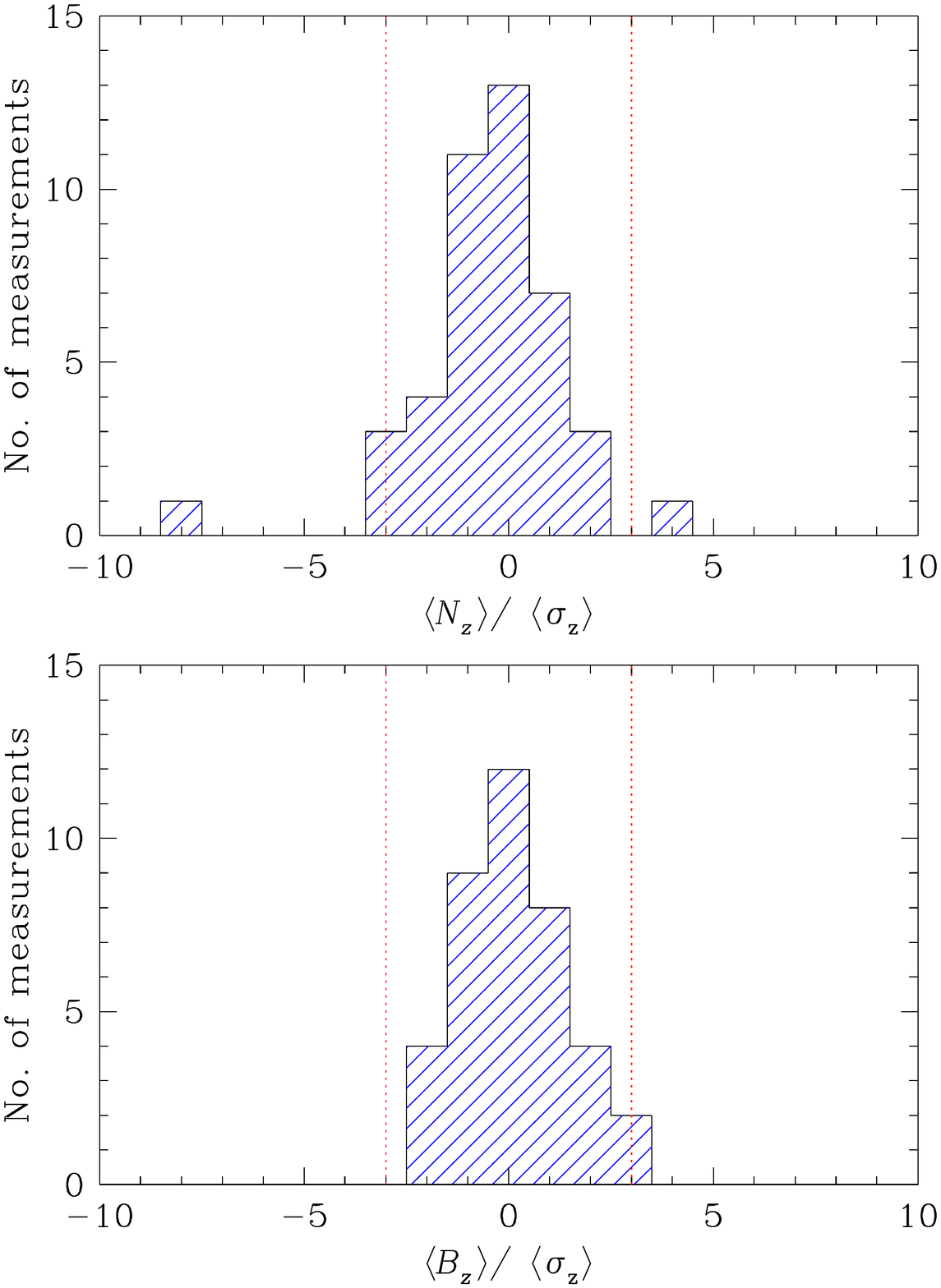}}
\caption{\label{Fig_Histonull} Histogram of the distribution of the null
  field values (top panel) and of the field values (bottom panel)
  normalised to their photon-noise error bars. Note that the field detections of
  the Galactic calibrators HD\,94660 and HD\,188041 are outside of the plot limits.}
\end{center}
\end{figure}

\subsubsection{AzV\,220}
Using the EW data published by \citet{Waletal15}, 
\citet{Nazetal15} could not find an unambiguous period from the
photometric data of AV\,220; however, a clear correlation between
photometric and spectroscopic variations was found by
\citet{Waletal15}.  Our new EWs are intermediate between the minimum
and maximum values, suggesting a small decline in recent
times. Simultaneous OGLE photometry also indicates some decrease
(M. Szymanski, priv.\ comm.), confirming the good correlation between
photometry and spectroscopy. However, this suggests that our
spectropolarimetry was not obtained close to the maximum of EW.

\subsubsection{2dFS\,936}\label{Sect_2dF_EW}
For this star we have considered the original B\&C spectra obtained in
October 2000 by \citeauthor{MasDuf01} (\citeyear{MasDuf01}; in their
Table~2 the star was labelled as ``Anon 1''), the data published
by \citet{Waletal15}, as well as the new data presented in Table~\ref{Tab_EWs},
namely: the measurements from our FORS2 data
obtained in October 2015, from a B\&C spectrum obtained in May 2016,
and from UVES-FLAMES spectra obtained in October 2015 and in November
2016. \citet{MasDuf01} reported only the EW of \HeII; from the
same spectra we also measured the EW of H$\beta$ ($-2.28\pm0.05$).
Our new 2015 FORS2 data were obtained at a very favorable phase
($\phi=0.4$), close to that expected for the maximum emission
($\phi=0.5$). The measured EWs however yield a surprise: the emissions
are much stronger than were seen up to now. The UVES-FLAMES spectrum
(see Fig.~\ref{Fig_Flames}) and our low-resolution spectrum taken in
May 2016 with the B\&C confirms this emission increase;
Figure~\ref{Fig_2dFS} shows the comparison between data obtained in
October 2016 and May 2016 with a previous spectrum obtained with the
B\&C in May 2012 \citep{Waletal15}. Also the spectrum obtained in
October 2000 by \citeauthor{MasDuf01} results in an EW value similar
to those obtained in 2016. While this \textit{a priori} enhances the chances
of detection, it is at odds with the expectation that Of?p have a
strictly repetitive behaviour. In addition, the recent photometry of
2dFS\,936 does indicate a slight brightening, but this was not the
case in October 2000 (M. Szymanski, priv.\ comm.).

\subsubsection{BI\,57}
\citet{Nazetal15}
suggested two photometric periods: 400\,d and 787\,d, with the latter
providing a better correlation with spectroscopic variations
\citep{Waletal15}. The new EW measurements are clearly at odds with
the 400\,d period, but perfectly in phase with the 787\,d period.
However, BI\,57 now appears to display two photometric maxima but only
one peak for the emission EWs, hence does not belong to the ``double
wave'' category of magnetic O-type stars, such as HD\,57682
\citep{Gruetal12} and CPD$-28^\circ$\,2561 \citep{Wadetal15}, which
have double maxima in both photometry and EWs.  This unexpected
behaviour has the consequence that our scheduling was not optimal:
BI\,57 was observed close to the EW minimum, hence probably when also
the longitudinal field is expected to have a small absolute value - if
the latter correlates with the emission line variations rather than
the photometry.

\subsubsection{SMC\,159-2}\label{Sect_EW_SMC}

For this star we have used the data from \citet{Masetal14},
\citep[also used by][]{Nazetal15}, and our new FORS2 measurements of
Table~\ref{Tab_EWs}. Note that \citet{Masetal14} and \citet{Nazetal15}
did not publish the EW measurements for H$\beta$ that we use in
Fig.~\ref{Fig_EWs} and that therefore we report here: $-2.76\pm0.10$\,\AA\
in 2013 and $-3.00\pm0.10$\,\AA\ in 2014 (emission component only).
For SMC\,159-2, our spectropolarimetric data were taken at a
rotational phase ($\phi$=0.49) very close to that of the maximum
emission (0.50) and the measured EWs are indeed close -- and slightly
higher than -- those measured in 2014 at $\phi$=0.6. Our new
spectropolarimetric data were indeed obtained at a very favourable
time for the measurements of the magnetic field.

\subsubsection{LMC\,164-2}
From the spectra obtained by \citet{Masetal14} at HJD=2456640.755
  ($\phi=0.29$) we measured the EW of the emission components of
  \HeII\ ($-1.80\pm0.10$\,\AA) and H$\beta$ ($-0.83\pm0.03$\,\AA).
We also note that for \HeII, \citet{Masetal14} incorrectly reported
the value of $\log(-{\rm EW}) = -0.2$ instead of  $\log(-{\rm EW}) = +0.2$;
therefore in this paper we considered the value of EW $-10^{0.2}=-1.58$.
Our new data were obtained at phases further from the
photometric maximum than the existing spectrum and,
accordingly, the new EWs indicate smaller emission strengths than
previously measured. Note however that the new data, even if not
perfectly scheduled, were still taken far from the minimum emission.

\section{Discussion}
\subsection{Quality check of field measurements:
  Galactic calibrators and statistical considerations}
Our field measurements of HD\,94660 (obtained on 2016-01-05 and on
2016-02-01) were found to be fully consistent with each other, and
consistent with the expected value of $\sim -2$\,kG (see
Sect.~\ref{Sect_FORS2_Spectropol}). Figure~~\ref{Fig_HD188041b}
shows that our measurement
of HD\,188041 (obtained on 2016-10-08) is in line with previous
FORS2 values, but higher than those obtained in the 50s and in the 90s
by \citet{Babcock54,Babcock58,Mathys94,MatHub97} and phased with the
rotation period given by \citet{LanMat00}. A misalignment of the retarder
waveplate would cause a decrease of the absolute value of the magnetic
field, and could possibly change its sign, but could not increase the
polarization signal \citep[see Eq.~55 of][]{Bagetal09}.  Vice versa, it
has been known for decades that the measured field strength does
depend on instrument and instrument setting \citep[e.g.][]{Henetal79},
and the case of FORS2 has been discussed in detail by
\citet{Lanetal14}. Therefore we conclude that all our measurements of
HD\,94660 and HD\,188041 confirm that the position of the FORS2
retarder waveplate was correctly reported by
the instrument encoders, and that it is unlikely that any field detection
was missed because of instrumental problems.

Null field values were found to be reasonably close to zero within
error bars. The top panel of Fig.~\ref{Fig_Histonull} shows the
histogram of the distribution $\nz/\sigma_z$, which, in the ideal
case, should be a Gaussian distribution centred about zero with
$\sigma = 1$.  Deviations from the ideal behaviour are not unexpected,
as photon-noise is not the only source of uncertainty in our
measurements. In particular, the remarkable outlier $\nz/\sigma_z =-8
$ comes from the field measurment of star HD\,188041, and simply
highlights the fact that when S/N is pushed to extremely high values,
other sources of noise become predominant, most likely tiny instrument
flexures and/or seeing variations, as discussed thoroughly by
\citet{Bagetal13}.  In addition, neighbouring stars show no detectable
field -- the two measurements in the $2.5 \le \bz / \sigma_z \le
3.5$ interval corresponds to the nearly 3\,$\sigma$ detection in
SMC\,159-2 and to the B0(V) star 2MASS J00500476-7321027.

The triple check (known magnetic stars, null diagnostics, neighbouring
objects) indicate the absence of problems in our data.

\subsection{Magnetic field measurements of Of?p stars}
Because of the exceptional strength of its H$\alpha$ emission
(compared to Galactic Of?p stars), we expected SMC\,159-2 to have an
especially strong field, with a dipolar strength $>10$\,kG
\citep{Nazetal15}. In fact, even such a strong field could escape
detection if observed with FORS2 at unfavourable geometrical conditions, for
instance, when a dipole field is seen with the dipolar axis
perpendicular to the line of sight. No matter the dipolar field strength
is, the average of the magnetic field component along the line
of sight would be zero.  However, as discussed in
Sect.~\ref{Sect_FORS2_Spectropol}, in Galactic Of?p stars, the
strength of the longitudinal field is correlated with the light curve
and the emission lines, i.e., the longitudinal field has a maximum
when the light curve and emission-line intensities have maxima, and
as discussed in Sect.~\ref{Sect_EW_SMC}, the spectropolarimetric data of
SMC\,159-2 were taken close to that phase.

In the star SMC\,159-2 we measured $\bz\ = 2.8 \pm 1$\,kG. This
  is {\it nearly} a $3\,\sigma$ detection of a 3\,kG longitudinal
  field, which is roughly the value that we would expect from a 10\,kG
  dipolar field observed pole-on. However, as extensively discussed by
  \citet{Bagetal12} and \citet{Bagetal13}, it is not possible to
  assign the classical statistical significance to the error bars
  formally derived from photon noise (that in this case would lead to
  the conclusion that a magnetic field has been detected with a very
  high confidence).  \citet{Bagetal12} have argued that because of
  tiny instrument instabilities, as well as the uncertainty introduced
  by the choices made during data reduction, a field measurement with
  FORS1/2 may be considered to be a firm detection only when it
  corresponds to at least 5\,$\sigma$ confidence (but note that
  $5\,\sigma$ is a general guideline and not a number derived from
  rigorous theoretical considerations). We should note that we also
  detected a null field at the $2\,\sigma$ level in the same spectrum,
  which might further weaken the credibility of our \bz\ detection at
  $2.8\,\sigma$ level. Therefore we conclude that our field detection
  is too marginal to be considered reliable. At the same time, we can
  assume that with a formal error bar of $\sim 1$\,kG, we would have
  certainly detected a 5\,kG longitudinal field, and we conclude that
  it is unlikely that SMC\,159-2, at the time of our observation,
  exhibited a longitudinal field $\ga 5$\,kG. We conclude that our
measurement of SMC\,159-2 is not sufficiently precise to fully rule
out a dipolar field of the strength predicted by \citet{Nazetal15},
but most probably SMC\,159-2 is not a star with magnetic properties
similar to the Galactic Of?p star NGC 1624-2, that is reported to have
a dipole field of nearly 20 kG.  More observations are needed to
better constrain the field of SMC\,159-2.

We have discovered that the star 2dFS\,936 also had a strong emission
in \HeII\ and H$\beta$ at the epoch of our observations. Although
H$\alpha$ was not as strong as that measured previously for SMC\,159-2
\citep[$-12.12$\,\AA\ against a maximum of $-19.2$\,\AA\ measured on
  SMC\,159-2 by][]{Nazetal15}, we still expect that 2dFS\,936 would
exhibit a strong and detectable field. However, for 2dFS\,936 we have
measured $\bz = -1\pm0.5$\,kG, which sets a lower limit to $\sim
2.0-2.5$\,kG to the actual \bz\ value.

\citet{Vinetal01} predicted $\dot{M}\,v_{\infty} \propto Z^{0.85}$,
where $\dot{M}\,v_{\infty}$ is the wind momentum and $Z$ the stellar
metallicity (see their Eq.~25). Given the lower metal content of the
SMC ($\sim 0.2\,Z_\sun$), we would expect a wind momentum a factor of
four smaller than in Galactic O stars ($0.2^{0.85} \sim
0.25$). Therefore, in the SMC, the wind confinement parameter
\[
\eta_\ast = \frac{B_\ast^2\,R_\ast^2}{\dot{M}\,v_{\infty}}
\]
(where $B_\ast$ is the field strength at the surface of the star with radius $R_\ast$)
reaches the same value for half the field strength needed in our
Galaxy.  This might explain why in the Magellanic Clouds we can detect
stars with strong emission lines and no strong magnetic
fields.  Nevertherless our lack of field detection in
2dFS\,936 somewhat weakens our hypothesis that the strength of the
emission lines and that of the longitudinal field may both be
explained in terms of the Oblique Rotator Model (ORM).

For the other three Of?p stars, no specific expectation (even
qualitative) on the magnetic strength could be made. Our data indicate
no detection for any of the stars: $0.5\pm0.6$\,kG in AzV\,220;
$-0.36\pm0.35$\,kG in BI\,57; $0.20 \pm 0.56$\,kG and $-0.55 \pm
0.90$\,kG in LMC\,164-2.  5$\sigma$ upper limits on $\vert\bz\vert$
amount to 3, 1.7, and 3\,kG for AzV\,220, LMC\,164-2, and BI\,57,
respectively. As seen in previous sections, data were not
obtained at an optimal time for AzV\,220 and LMC\,164-2, and even far
from it for BI\,57, limiting our detection capability.

\citet{Wade15} summarised that the inferred surface dipole
  strengths of magnetic O stars (including Of?p stars) are typically in
  the range 1--3\,kG, \citep[see Fig.~4 of][]{Wade15}. The smallest dipolar strength inferred in an
Of?p star is of a few hundred G (an upper limit estimated for
HD\,37742 by Blaz\`ere et al., in preparation), followed by $\la
1$\,kG in the case of HD\,148937 \citep{Wadetal12a}, to $\sim$\,2.5kG
in HD\,191612 \citep[discovered as Of?p star by][]{Walborn73}, and in
CPD\,$-28^{\circ}$2561. However, note that HD\,191612 and
CPD\,$-28^{\circ}$2561 show at most a longitudinal field of
$\vert\bz\vert \sim 600$\,G \citep{Wadetal11,Wadetal15}. Stronger
field seem relatively rare, the most notable exception being NGC\,1624-2
with a dipolar field strength of $\sim 20$\,kG
\citep{Wadetal12b}.  Therefore, the main conclusion that can be drawn
so far is that the magnetic fields of Of?p stars in the Magellanic Clouds are not
much stronger than in our Galaxy.

\section{Conclusions}

We have analysed the five known Of?p stars in the Magellanic Clouds
with spectropolarimetric techniques to search for evidence of their
magnetic fields.  Checks using standard stars, null profiles, and
neighbouring stars (not expected to be magnetic) were performed, and
validated the reliability of our results.  No magnetic field was
detected, though some of our detection limits were higher than
  expected due to bad weather and (perhaps) to the fact that the
epoch of observations did not always correspond to the expected phase
of field maximum. This led to poorer upper limits on any undetected
magnetic field that might be present.

Our expectations was that SMC\,159-2, which was observed during a
period of strong emission, would exhibit a strong magnetic field,
possibly as large as 10\,kG. Our data rule out the presence of
longitudinal fields stronger than 5\,kG (which admittedly are very
rare amongst Galactic magnetic stars). For 2dFS\,936, which also
showed unexpectedly very strong emission lines, our data rule out a
longitudinal field stronger than 2.0-2.5\,kG at the epoch of our
observations. The lack of a detection of a strong field in a star with
such very strong emission lines is at odd with our expectations.

The (non-polarimetric) spectroscopic observations of SMC\,159-2
  and LMC\,164-2 are fully consistent with the previously observed behaviour
  of these stars, and those of AzV\,220 indicate a recent, small
  decline of the emission line strength. In the remaining two
    stars, our spectroscopic observations reveal some unexpected
  features. {\it i)} They bring additional evidence that the best
  period of BI\,57 is likely 787\,d, showing that this star presents
  two photometric peaks for one peak in emission line strength, a
  feature never observed up to now in Galactic magnetic O-type stars. {\it
    ii)} They show a very strong and unexpected strengthening
  of the emission lines of 2dFS\,936, suggesting that the behaviour
    of this star may not be as reproducible as for other Of?p stars.

Our findings suggest that the nature of Of?p stars may
  not be fully understood in terms of the Oblique Rotator Model, and
  call for a closer monitoring to better
understand the interplay between magnetic fields and stellar
winds, and to investigate the role of stellar metallicity in the
magnetospheres of Of?p stars.  More specifically, further photometric,
spectroscopic and spectropolarimetric monitoring in order to better
sample the rotational cycle (especially of BI\,57, SMC\,159-2 and
  2dFS\,936) could set more stringent constraints on the modelling of
the circumstellar environments of massive stars.

\begin{acknowledgements}
  Based on observations made with ESO Telescopes at the La
  Silla-Paranal Observatory under programme IDs 096.D-0856 (FORS
  specropolarimetric survey) and 096.D-0604(A) e 098.D-0876(A)
  (FLAMES spectra). We also
  acknowledge the use of archive data from programme ID 087.D-0870.
  YN acknowledges support from the Fonds National de la Recherche
  Scientifique (Belgium) and the University of Liege, especially for
  travel to Chile, as well as general support from the PRODEX XMM and
  Integral contracts and the `Action de Recherche Concert\'ee' (CFWB).
  GAW acknowledges Discovery Grant support from he Natural Science and
  Engineering Research Council (NSERC) of Canada.  We thank Michal
  Szymanski for making us available some preliminary results from OGLE
  photometry, and G\"{o}tz Gr\"{a}fener for useful discussions.
\end{acknowledgements}
\bibliography{sbabib}

\begin{thebibliography}{47}
\expandafter\ifx\csname natexlab\endcsname\relax\def\natexlab#1{#1}\fi

\bibitem[{{Appenzeller} {et~al.}(1998){Appenzeller}, {Fricke}, {F{\"u}rtig},
  {G{\"a}ssler}, {H{\"a}fner}, {Harke}, {Hess}, {Hummel}, {J{\"u}rgens},
  {Kudritzki}, {Mantel}, {Meisl}, {Muschielok}, {Nicklas}, {Rupprecht},
  {Seifert}, {Stahl}, {Szeifert}, \& {Tarantik}}]{Appetal98}
{Appenzeller}, I., {Fricke}, K., {F{\"u}rtig}, W., {et~al.} 1998, The
  Messenger, 94, 1

\bibitem[{{Appenzeller} \& {Rupprecht}(1992)}]{AppRup92}
{Appenzeller}, I. \& {Rupprecht}, G. 1992, The Messenger, 67, 18

\bibitem[{{Babcock}(1954)}]{Babcock54}
{Babcock}, H.~W. 1954, \apj, 120, 66

\bibitem[{{Babcock}(1958)}]{Babcock58}
{Babcock}, H.~W. 1958, \apjs, 3, 141

\bibitem[{{Bagnulo} {et~al.}(2013){Bagnulo}, {Fossati}, {Kochukhov}, \&
  {Landstreet}}]{Bagetal13}
{Bagnulo}, S., {Fossati}, L., {Kochukhov}, O., \& {Landstreet}, J.~D. 2013,
  \aap, 559, A103

\bibitem[{{Bagnulo} {et~al.}(2015){Bagnulo}, {Fossati}, {Landstreet}, \&
  {Izzo}}]{Bagetal15}
{Bagnulo}, S., {Fossati}, L., {Landstreet}, J.~D., \& {Izzo}, C. 2015, \aap,
  583, A115

\bibitem[{{Bagnulo} {et~al.}(2009){Bagnulo}, {Landolfi}, {Landstreet}, {Landi
  Degl'Innocenti}, {Fossati}, \& {Sterzik}}]{Bagetal09}
{Bagnulo}, S., {Landolfi}, M., {Landstreet}, J.~D., {et~al.} 2009, \pasp, 121,
  993

\bibitem[{{Bagnulo} {et~al.}(2012){Bagnulo}, {Landstreet}, {Fossati}, \&
  {Kochukhov}}]{Bagetal12}
{Bagnulo}, S., {Landstreet}, J.~D., {Fossati}, L., \& {Kochukhov}, O. 2012,
  \aap, 538, A129

\bibitem[{{Bagnulo} {et~al.}(2006){Bagnulo}, {Landstreet}, {Mason}, {Andretta},
  {Silaj}, \& {Wade}}]{Bagetal06}
{Bagnulo}, S., {Landstreet}, J.~D., {Mason}, E., {et~al.} 2006, \aap, 450, 777

\bibitem[{{Bagnulo} {et~al.}(2002){Bagnulo}, {Szeifert}, {Wade}, {Landstreet},
  \& {Mathys}}]{Bagetal02}
{Bagnulo}, S., {Szeifert}, T., {Wade}, G.~A., {Landstreet}, J.~D., \& {Mathys},
  G. 2002, \aap, 389, 191

\bibitem[{{Casini} \& {Landi Degl'Innocenti}(1994)}]{CasLan94}
{Casini}, R. \& {Landi Degl'Innocenti}, E. 1994, \aap, 291, 668

\bibitem[{{Donati} {et~al.}(2002){Donati}, {Babel}, {Harries}, {Howarth},
  {Petit}, \& {Semel}}]{Donetal02}
{Donati}, J.-F., {Babel}, J., {Harries}, T.~J., {et~al.} 2002, \mnras, 333, 55

\bibitem[{{Donati} {et~al.}(2006){Donati}, {Howarth}, {Bouret}, {Petit},
  {Catala}, \& {Landstreet}}]{Donetal06}
{Donati}, J.-F., {Howarth}, I.~D., {Bouret}, J.-C., {et~al.} 2006, \mnras, 365,
  L6

\bibitem[{{Donati} \& {Landstreet}(2009)}]{DonLan09}
{Donati}, J.-F. \& {Landstreet}, J.~D. 2009, \araa, 47, 333

\bibitem[{{Ferrario} {et~al.}(2009){Ferrario}, {Pringle}, {Tout}, \&
  {Wickramasinghe}}]{Feretal09}
{Ferrario}, L., {Pringle}, J.~E., {Tout}, C.~A., \& {Wickramasinghe}, D.~T.
  2009, \mnras, 400, L71

\bibitem[{{Fossati} {et~al.}(2015){Fossati}, {Castro}, {Sch{\"o}ller},
  {Hubrig}, {Langer}, {Morel}, {Briquet}, {Herrero}, {Przybilla}, {Sana},
  {Schneider}, {de Koter}, \& {BOB Collaboration}}]{Fosetal15}
{Fossati}, L., {Castro}, N., {Sch{\"o}ller}, M., {et~al.} 2015, \aap, 582, A45

\bibitem[{{Gr{\"a}fener} {et~al.}(2012){Gr{\"a}fener}, {Owocki}, \&
  {Vink}}]{Graetal12}
{Gr{\"a}fener}, G., {Owocki}, S.~P., \& {Vink}, J.~S. 2012, \aap, 538, A40

\bibitem[{{Grunhut} {et~al.}(2017){Grunhut}, {Wade}, {Neiner}, {Oksala},
  {Petit}, {Alecian}, {Bohlender}, {Bouret}, {Henrichs}, {Hussain},
  {Kochukhov}, \& {MiMeS Collaboration}}]{Gruetal17}
{Grunhut}, J.~H., {Wade}, G.~A., {Neiner}, C., {et~al.} 2017, \mnras, 465, 2432

\bibitem[{{Grunhut} {et~al.}(2012){Grunhut}, {Wade}, {Sundqvist}, {ud-Doula},
  {Neiner}, {Ignace}, {Marcolino}, {Rivinius}, {Fullerton}, {Kaper},
  {Mauclaire}, {Buil}, {Garrel}, {Ribeiro}, \& {Ubaud}}]{Gruetal12}
{Grunhut}, J.~H., {Wade}, G.~A., {Sundqvist}, J.~O., {et~al.} 2012, \mnras,
  426, 2208

\bibitem[{{Hensberge} {et~al.}(1979){Hensberge}, {van Rensbergen}, {Deridder},
  \& {Goossens}}]{Henetal79}
{Hensberge}, H., {van Rensbergen}, W., {Deridder}, G., \& {Goossens}, M. 1979,
  \aap, 75, 83

\bibitem[{{Landstreet} {et~al.}(2014){Landstreet}, {Bagnulo}, \&
  {Fossati}}]{Lanetal14}
{Landstreet}, J.~D., {Bagnulo}, S., \& {Fossati}, L. 2014, \aap, 572, A113

\bibitem[{{Landstreet} \& {Mathys}(2000)}]{LanMat00}
{Landstreet}, J.~D. \& {Mathys}, G. 2000, \aap, 359, 213

\bibitem[{{Langer}(2012)}]{Langer12}
{Langer}, N. 2012, \araa, 50, 107

\bibitem[{{Maitzen} {et~al.}(2001){Maitzen}, {Paunzen}, \&
  {Pintado}}]{Maietal01}
{Maitzen}, H.~M., {Paunzen}, E., \& {Pintado}, O.~I. 2001, \aap, 371, L5

\bibitem[{{Massey} \& {Duffy}(2001)}]{MasDuf01}
{Massey}, P. \& {Duffy}, A.~S. 2001, \apj, 550, 713

\bibitem[{{Massey} {et~al.}(2014){Massey}, {Neugent}, {Morrell}, \&
  {Hillier}}]{Masetal14}
{Massey}, P., {Neugent}, K.~F., {Morrell}, N., \& {Hillier}, D.~J. 2014, \apj,
  788, 83

\bibitem[{{Mathys}(1994)}]{Mathys94}
{Mathys}, G. 1994, \aaps, 108

\bibitem[{{Mathys} \& {Hubrig}(1997)}]{MatHub97}
{Mathys}, G. \& {Hubrig}, S. 1997, \aaps, 124

\bibitem[{{Mokiem} {et~al.}(2007){Mokiem}, {de Koter}, {Vink}, {Puls}, {Evans},
  {Smartt}, {Crowther}, {Herrero}, {Langer}, {Lennon}, {Najarro}, \&
  {Villamariz}}]{Moketal07}
{Mokiem}, M.~R., {de Koter}, A., {Vink}, J.~S., {et~al.} 2007, \aap, 473, 603

\bibitem[{{Moss}(2001)}]{Moss01}
{Moss}, C. 2001, in Astronomical Society of the Pacific Conference Series, Vol.
  230, Galaxy Disks and Disk Galaxies, ed. J.~G. {Funes} \& E.~M. {Corsini},
  487--490

\bibitem[{{Naz{\'e}} {et~al.}(2012){Naz{\'e}}, {Bagnulo}, {Petit}, {Rivinius},
  {Wade}, {Rauw}, \& {Gagn{\'e}}}]{Nazetal12}
{Naz{\'e}}, Y., {Bagnulo}, S., {Petit}, V., {et~al.} 2012, \mnras, 423, 3413

\bibitem[{{Naz{\'e}} {et~al.}(2015){Naz{\'e}}, {Walborn}, {Morrell}, {Wade}, \&
  {Szyma{\'n}ski}}]{Nazetal15}
{Naz{\'e}}, Y., {Walborn}, N.~R., {Morrell}, N., {Wade}, G.~A., \&
  {Szyma{\'n}ski}, M.~K. 2015, \aap, 577, A107

\bibitem[{{Paunzen} {et~al.}(2011){Paunzen}, {Netopil}, \& {Bord}}]{Pauetal11}
{Paunzen}, E., {Netopil}, M., \& {Bord}, D.~J. 2011, \mnras, 411, 260

\bibitem[{{Sanyal} {et~al.}(2017){Sanyal}, {Langer}, {Sz{\'e}csi}, {-C Yoon},
  \& {Grassitelli}}]{Sanetal17}
{Sanyal}, D., {Langer}, N., {Sz{\'e}csi}, D., {-C Yoon}, S., \& {Grassitelli},
  L. 2017, \aap, 597, A71

\bibitem[{{Stibbs}(1950)}]{Stibbs50}
{Stibbs}, D.~W.~N. 1950, \mnras, 110, 395

\bibitem[{{Sundqvist} {et~al.}(2012){Sundqvist}, {ud-Doula}, {Owocki},
  {Townsend}, {Howarth}, \& {Wade}}]{Sunetal12}
{Sundqvist}, J.~O., {ud-Doula}, A., {Owocki}, S.~P., {et~al.} 2012, \mnras,
  423, L21

\bibitem[{{Vink} {et~al.}(2001){Vink}, {de Koter}, \& {Lamers}}]{Vinetal01}
{Vink}, J.~S., {de Koter}, A., \& {Lamers}, H.~J.~G.~L.~M. 2001, \aap, 369, 574

\bibitem[{{Wade}(2015)}]{Wade15}
{Wade}, G.~A. 2015, in Astronomical Society of the Pacific Conference Series,
  Vol. 494, Physics and Evolution of Magnetic and Related Stars, ed. Y.~Y.
  {Balega}, I.~I. {Romanyuk}, \& D.~O. {Kudryavtsev}, 30

\bibitem[{{Wade} {et~al.}(2015){Wade}, {Barb{\'a}}, {Grunhut}, {Martins},
  {Petit}, {Sundqvist}, {Townsend}, {Walborn}, {Alecian}, {Alfaro},
  {Ma{\'{\i}}z Apell{\'a}niz}, {Arias}, {Gamen}, {Morrell}, {Naz{\'e}}, {Sota},
  {ud-Doula}, \& {MiMeS Collaboration}}]{Wadetal15}
{Wade}, G.~A., {Barb{\'a}}, R.~H., {Grunhut}, J., {et~al.} 2015, \mnras, 447,
  2551

\bibitem[{{Wade} {et~al.}(2012{\natexlab{a}}){Wade}, {Grunhut}, {Gr{\"a}fener},
  {Howarth}, {Martins}, {Petit}, {Vink}, {Bagnulo}, {Folsom}, {Naz{\'e}},
  {Walborn}, {Townsend}, \& {Evans}}]{Wadetal12b}
{Wade}, G.~A., {Grunhut}, J., {Gr{\"a}fener}, G., {et~al.} 2012{\natexlab{a}},
  \mnras, 419, 2459

\bibitem[{{Wade} {et~al.}(2011){Wade}, {Howarth}, {Townsend}, {Grunhut},
  {Shultz}, {Bouret}, {Fullerton}, {Marcolino}, {Martins}, {Naz{\'e}}, {Ud
  Doula}, {Walborn}, \& {Donati}}]{Wadetal11}
{Wade}, G.~A., {Howarth}, I.~D., {Townsend}, R.~H.~D., {et~al.} 2011, \mnras,
  416, 3160

\bibitem[{{Wade} {et~al.}(2012{\natexlab{b}}){Wade}, {Ma{\'{\i}}z
  Apell{\'a}niz}, {Martins}, {Petit}, {Grunhut}, {Walborn}, {Barb{\'a}},
  {Gagn{\'e}}, {Garc{\'{\i}}a-Melendo}, {Jose}, {Moffat}, {Naz{\'e}}, {Neiner},
  {Pellerin}, {Penad{\'e}s Ordaz}, {Shultz}, {Sim{\'o}n-D{\'{\i}}az}, \&
  {Sota}}]{Wadetal12a}
{Wade}, G.~A., {Ma{\'{\i}}z Apell{\'a}niz}, J., {Martins}, F., {et~al.}
  2012{\natexlab{b}}, \mnras, 425, 1278

\bibitem[{{Wade} {et~al.}(2016){Wade}, {Neiner}, {Alecian}, {Grunhut}, {Petit},
  {Batz}, {Bohlender}, {Cohen}, {Henrichs}, {Kochukhov}, {Landstreet},
  {Manset}, {Martins}, {Mathis}, {Oksala}, {Owocki}, {Rivinius}, {Shultz},
  {Sundqvist}, {Townsend}, {ud-Doula}, {Bouret}, {Braithwaite}, {Briquet},
  {Carciofi}, {David-Uraz}, {Folsom}, {Fullerton}, {Leroy}, {Marcolino},
  {Moffat}, {Naz{\'e}}, {Louis}, {Auri{\`e}re}, {Bagnulo}, {Bailey},
  {Barb{\'a}}, {Blaz{\`e}re}, {B{\"o}hm}, {Catala}, {Donati}, {Ferrario},
  {Harrington}, {Howarth}, {Ignace}, {Kaper}, {L{\"u}ftinger}, {Prinja},
  {Vink}, {Weiss}, \& {Yakunin}}]{Wadetal16}
{Wade}, G.~A., {Neiner}, C., {Alecian}, E., {et~al.} 2016, \mnras, 456, 2

\bibitem[{{Walborn}(1972)}]{Walborn72}
{Walborn}, N.~R. 1972, \aj, 77, 312

\bibitem[{{Walborn}(1973)}]{Walborn73}
{Walborn}, N.~R. 1973, \aj, 78, 1067

\bibitem[{{Walborn} {et~al.}(2011){Walborn}, {Ma{\'{\i}}z Apell{\'a}niz},
  {Sota}, {Alfaro}, {Morrell}, {Barb{\'a}}, {Arias}, \& {Gamen}}]{Waletal11}
{Walborn}, N.~R., {Ma{\'{\i}}z Apell{\'a}niz}, J., {Sota}, A., {et~al.} 2011,
  \aj, 142, 150

\bibitem[{{Walborn} {et~al.}(2015){Walborn}, {Morrell}, {Naz{\'e}}, {Wade},
  {Bagnulo}, {Barb{\'a}}, {Ma{\'{\i}}z Apell{\'a}niz}, {Howarth}, {Evans}, \&
  {Sota}}]{Waletal15}
{Walborn}, N.~R., {Morrell}, N.~I., {Naz{\'e}}, Y., {et~al.} 2015, \aj, 150, 99

\end{thebibliography}
\end{document}